# A hybrid cross entropy algorithm for solving dynamic transit network design problem

TAI-YU MA

*Transport Economics Laboratory*
*University Lyon 2 - CNRS*
*Lyon, France*

This paper proposes a hybrid multiagent learning algorithm for solving the dynamic simulation-based bilevel network design problem. The objective is to determine the optimal frequency of a multimodal transit network, which minimizes total users' travel cost and operation cost of transit lines. The problem is formulated as a bilevel programming problem with equilibrium constraints describing non-cooperative Nash equilibrium in a dynamic simulation-based transit assignment context. A hybrid algorithm combing the cross entropy multiagent learning algorithm and Hooke-Jeeves algorithm is proposed. Computational results are provided on the Sioux Falls network to illustrate the performance of the proposed algorithm.

## 1. INTRODUCTION

Transit network design problem (TNDP) has been an important problem in transportation science and widely studied in the past [1][2][3]. The objective is to determine optimal transit line frequencies in a transit network, which minimizes total user cost and transit operation cost under resource and user equilibrium flow constraints. The later results from the solution of traffic assignment problem aiming to determine the traffic flow at Nash equilibrium (user optimal) states. The TNDP can be generally formulated as a bilevel programming problem, where the upper-level is a constrained minimization problem for the optimal transit line frequency decision; the lower-level is a variational inequality (VI) problem for solving the user-optimal equilibrium flow. As the evaluation of the objective function at the upper-level requires a solution of the VI problem at the lower-level, the problem has been well known as a difficult problem in mathematical programming and transportation science.

In the past, the TNDP has been studied by many authors. Gao et al. [4] formulated the TNDP as a bilevel programming model and proposed a solution procedure based on sensitivity analysis. The upper-level problem is formulated as a minimization problem under the equilibrium transit assignment constraints. The lower-level problem is formulated as the VI problem with differential cost functions. The proposed approach requires the calculation of the derivatives of flow with respect to the line frequency to obtain optimal solutions. Marcotte [1] proposed a formal description of the TNDP and provided several heuristic procedures for solving it. The user-optimal flow is obtained by solving the VI problem in a static network with asymmetric link cost functions. LeBlanc proposed a series of papers for the TNDP [2][5]. In [2], a bilevel static multimodal transit network design model has been proposed. The author first solved a mode-split assignment problem to obtain a user-optimal equilibrium flow by Frank-Wolfe algorithm and then applied Hooke-Jeeves algorithm to iteratively derive optimal frequencies in a static transit network. Other solution techniques for the bilevel programming problem can be found in [6]. However, for dynamic simulation-based transit assignment, the above derivative-based methods cannot be applied since the functional form of the derivatives is generally unavailable. The simulation-based VI problem is generally difficult to solve in the dynamic transit system. For this issue, Ma and Lebacque [7][8] proposed a cross entropy (CE) based solution algorithm to iteratively derive optimal travel choice probabilities towards user equilibrium based on minimizing the Kullback-Liebler relative entropy (cross entropy) between two consecutive probability distributions.

In this work, a hybrid algorithm is proposed by combing the multiagent cross entropy learning algorithm and the Hooke-Jeeves algorithm for solving the simulation-based transit network design problem. The proposed algorithm is derivative-free, convenient for solving the simulation-based TNDP. For the transit system simulation, a multiagent approach is proposed to capture explicitly the transit system dynamics. We propose a multi-layer network to effectively represent the transit network and simulate the movement of different agents (passengers and vehicles). Passenger's waiting time at stop is explicitly calculated subject to the capacity constraint of the vehicle.

The rest of the paper is organized as follows. Section 2 describes the mathematical formulation of the bilevel programming problem for the TNDP. It follows in Section 3 the dynamic transit system description based on the multiagent approach along with the transit network model and travel cost formulation. Section 4 presents the proposed solution algorithm by combining the Hooke-Jeeves algorithm and the CE multiagent approach. A state-of-the-art algorithm based on the method of successive average (MSA) for solving simulation-based dynamic traffic assignment problem is proposed. Section 5 provides the computational results of the CE multagent approach and the MSA appraoch on the Sioux Falls network [2] to validate the obtained lower-level user euilibrium solution. Then we show the optimal transit frequency obtained by the hybrid algorithm. Section 6 concludes the paper.

## 2. THE TRANSIT NETWORK DESIGN MODEL

Notation
- $l$      transit line
- $L$      set of transit lines
- $\overline{Y}_l$      upper bound of the frequency of transit line $l$
- $\underline{Y}_l$      lower bound of the frequency of transit line $l$
- $y_l$      frequency of transit line $l$
- $\mathbf{y}$      vector of the frequency of transit lines
- $\theta_l$      cost increase for frequency in transit line $l$
- $m$      designation of a user
- $C_m$      generalized travel cost of user $m$
- $k$      origin-destination pair

| | |
|---|---|
| $K$ | set of origin-destination (o-d) pair $k$ |
| $f_r(t)$ | flow on path $r$ at time $t$ |
| $\mathbf{f}$ | vector of flows |
| $r$ | path index |
| $R_k$ | set of paths connecting o-d pair $k$ |
| $d_k(t)$ | demand of origin-destination pair at time $t$ |
| $D_k$ | demand of origin-destination pair |
| $t$ | time index |
| $T$ | the time of the last vehicle/user leaves the network |

The TNDP is formulated as a bilevel programming problem. For the upper-level problem, a decision maker aims to minimize the total cost of the transit system under feasible frequency constraints and dynamic user equilibrium at the lower-level. For the lower-level problem, each user aims to minimize his/her travel cost, this is a problem of noncooperative Nash equilibrium in a multiagent system. At the upper-level, the problem is formulated as a constrained minimization problem subject to feasible line frequency and to the equilibrium path flow determined at the lower-level. At the lower-level, the problem is formulated as a VI problem with the line frequencies imposed at the upper-level.

**(Upper-level)**

$$\text{Min} \quad Z(\mathbf{y},\mathbf{f}^*) = \sum_k \sum_{m \in d_k} C_m(t,\mathbf{f}^*) + \sum_l y_l \theta_l \tag{1}$$

$$\text{St.} \quad \underline{Y}_l \leq y_l \leq \overline{Y}_l, \forall l \in L \tag{2}$$

$\mathbf{f}^*$ is the noncooperative Nash equilibrium path flow vector determined by solving the minimization problem of (4)-(7).

(3)

**(Lower-level, VI problem)**

Find user optimal equilibrium path flow vector $\mathbf{f}$ such that

$$\text{Min} \quad S(\mathbf{f}) = \sum_k \sum_{r,s \in R_k} \int_{t_0}^T f_r [C_r(t,\mathbf{f}) - C_s(t,\mathbf{f})]_+ \tag{4}$$

$$\text{St.} \quad \sum_{r \in R_k} f_r(t) = d_k(t), \forall k \in K, \forall t \in [t_0, T] \tag{5}$$

$$\int_{t_0}^T d_k(t) = D_k, \quad \forall k \tag{6}$$

$$f_r(t) \geq 0, \forall r, \forall t \in [t_0, T], \tag{7}$$

where the function $[q]_+ = \max(0, q)$.

The objective function (1) minimizes total generalized travel cost of users and total operation cost. The constraints (2) mean that the line frequencies are bounded. The constraint (3) is the user optimal path flow vector satisfying the equilibrium condition

solved by (4)-(7) [9]. At the lower-level, the user equilibrium flow is stated as

$$f_r[C_r(t,\mathbf{f}) - C_s(t,\mathbf{f})]_+ = 0, \quad \forall r,s \in R_k, \forall k, \forall t \in [t_0, T] \tag{8}$$

can be obtianed by solving the minimization problem of (4)-(7). The user equilibrium flow states that for users of the same origin and destination (OD) the generalized travel cost resulting from departure time and route choices is equal and no less than that of unused choice alternatives. The constraints (5)-(7) state the conservation of flow and non-negativity of path flow.

Previous studies [10][11] showed that we can define a relative gap function to measure how the generalized travel cost is far from the idealized shortest path cost. The gap function is defiend as

$$\text{Gap}(\mathbf{f}) = \frac{\sum_{h \in H} \sum_{k \in K} \sum_{r \in R_{hk}} f_{hkr}[C_{hkr} - C_{hk}^*]}{\sum_{h \in H} \sum_{k \in K} d_{hk} C_{hk}^*}, \tag{9}$$

where

$h$ : departure time choice index $\forall h \in H$ with $H = \{0,1,2,...,n\}$. Given a selected departure time interval $h$, a random departure time will be taken within $[t_0 + h\Delta, t_0 + (h+1)\Delta)$ with $t_0$ the earliest departure time and $\Delta$ a small discretized time interval (e.g. 5 minuites).

$C_{hkr}$ : experienced travel cost with respect to ($h$, $k$, $r$)

$d_{hk}$ : time-dependent travel demand with respect to ($h$, $k$)

$C_{hk}^*$ : minimum path generalized travel cost with respect to ($h$, $k$)

The gap function reports the average gap towards to dynamic user equilibrium. When the gap function converges to a stable value and the obtained average travel cost for utilized departure time intervals and paths are no more than that on unused alternatives, the approximate of user equilibrium is achieved.

## 3. MULTIAGENT-BASED TRANSIT SYSTEM

To capture the dynamics of the movements of the users and the effect of congestion at stations, the multiagent approach is adopted. The multiagent approach is very convenient for simulating the dynamics of operations of transit vehicles and user flow on the system [12]. Its advantage resides on its flexibility in capturing the interactions between agents with their environment. We utilize a multilayer network structure to explicitly model complex connections within multimodal stations with the presence of different transport modes and service lines. The detail of the transit network and multiagent simulation is described as follows.

**3.1 Transit network and transit paths**

The transit network is represented by a directed graph $G(N, A)$, where $N$ is the set of nodes and $A$ the set of arcs. The nodes are classified into three types: origin/destination, station, and line node [13]. The network structure is illustrated in Figure 1. As shown in Fig. 1, the origin/destination nodes are connected with related serviced station nodes by walking arcs. The station nodes are connected with its service transit lines, with other station nodes within the same multimodal station and with origin/destination nodes. Each arc is characterized by its travel time calculated as its length divided by walking or constant mode-specific vehicle speed. A *transit path* (called *path* hereafter) is an acyclic path connecting an origin-destination (OD) pair in the multilayer transit network. The travel time of a path comprises walking time accessing to O/D, transit line nodes by boarding/alighting arcs, waiting time at line nodes, and transfer time between stations within the same multimodal station. The travel time on the boarding arcs represents average walking time from the station center to the boarding point of the vehicle. Hence, the *generalized travel cost* function for the path consists of the following parts: (a) walking time $\pi_r^w$; (b) in-vehicle time $\pi_r^v$; (c) waiting time $\pi_r^s(t,\mathbf{f})$; (d) mode transfer penalty $\lambda$; (e) early/late arrival penalty $\rho_r(t,\mathbf{f})$; (f) fare $\Gamma_r$. By assuming the First-In-First-Out principle for boarding a vehicle [14], the waiting time depends on the supply (vehicle capacity and service frequency) and the demand at each transit line node. Hence, the waiting time $\pi_i(t)$ for a user arriving at line node $i$ at time $t$ is calculated as $\pi_i(t) = D_i^{-1}(S_i(t)) - t$, where $S_i(t)$ is cumulative arrivals at line node $i$ by time $t$, $D_i^{-1}(t)$ is the inverse function of cumulative departure from line node $i$ by time $t$.

The generalized travel cost of path $r$ when departing from origin at time $t$ is then evaluated as:

$$C_r(t,\mathbf{f}) = \alpha(\pi_r^w + \pi_r^v + \pi_r^s(t,\mathbf{f})) + n_r\lambda + \rho_r(t,\mathbf{f}) + \Gamma_r, \qquad (10)$$

where $\alpha$ is the unitary monetary value of travel time obtained by travel survey data; $n_r$ is the number of mode change which can be directly calculated by the used multimodal path; $\lambda$ is the unitary penalty per change obtained by travel survey data; $\Gamma_r$ is fare of path $r$.

Based on the experimental study of Small [15], the early/late arrival penalty when arriving at destination at time $t^{arr}$ is defined as:

$$\rho_r(t,\mathbf{f}) = \mu_a \times \max(0, \tau - \varpi - t^{arr}) + \mu_b \times \max(0, t^{arr} - \tau - \varpi), \qquad (11)$$

where $\mu_a$ and $\mu_b$ are unitary penalty associated with early and late arrival, respectively. The value of unitary penalty can be generally obtained by travel behavior survey. $\tau$ is the desired arrival time at destination which is set as identical for simplicity; $\varpi$ is the half of tolerable schedule delay interval without penalty, generally set as 2-5 minutes.

### 3.2 Multiagent-based transit system

The system is composed of two classes of agents, i.e. transit vehicles and users. For the vehicle agent, it represents a mode-specific vehicle such as tramway/metro/train operating on respective transit lines with predefined frequency and capacity constraints. For simplicity, the vehicle agents move with constant speed neglecting accidents or delayed situation. The vehicle capacity is assumed fixed and transport mode-specific. When the vehicle agent arrives to a line node, a constant stop time makes the users board the vehicle. If the vehicle capacity is achieved, not serviced users need to wait for the next vehicle at the same line node.

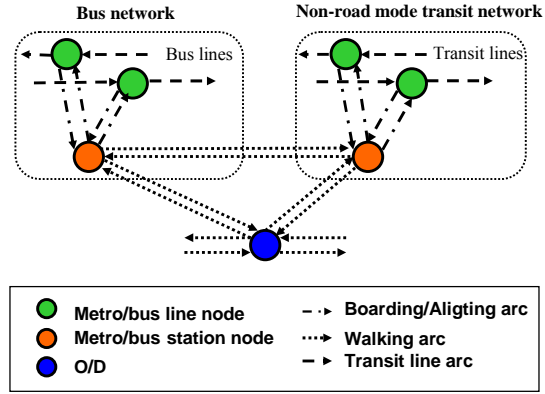

Figure 1  Transit t network structure

For the user agent, each one iteratively adapts his/her departure time and path choice in order to minimize his/her generalized travel cost. The user behavior is based on the bounded-rationality assumption [16], assuming that the users have no complete information about the travel choice decision of the other users, neither the real-time congestion information of the transit network. The departure time and path choices are adjusted in a day-to-day basis based on the performance of choice alternatives on the previous day. The learning process is similar to the reinforcement learning process where users shift their choices to more attractive alternatives based on past experiences. The difficulty is how to update the choice probability towards the user equilibrium in a multiagent system.

## 4. SOLUTION ALGORITHM

In this section, a hybrid solution algorithm is proposed for solving the dynamic simulation-based TNDP. As user's experienced travel cost depends on transit network supply (vehicle capacity and line frequency) and travel demand dynamics (users of the same OD competing the same resources), the derivative-based methods cannot be applied to solve the proposed bilevel programming problem. The algorithm is composed of two main steps. First, the cross entropy learning algorithm is applied for solving the lower-level VI problem. The solution obtained at this stage is then utilized for the computation of the value of the objective function at the upper-level. Secondly, to optimize the line frequencies, the Hooke-Jeeves algorithm is applied, which iteratively finds the optimal configuration of line frequency by moving each line frequency towards to better solutions [17][2]. The solution algorithms are described as follows.

### 4.1 Hooke-Jeeves algorithm

**Step 1:** Initialize the vector $\mathbf{y}_0$ of line frequency, set exploratory search step size as $\Delta = \Delta_0$. Set iteration index i=0.

**Step 2:** For each $l \in \underline{L}$, set trial frequency vector $\mathbf{y}'$ by increasing $y_l$ in $\mathbf{y}_i$ by $\Delta$, if $y_l + \Delta > \overline{Y}_l$ set $y_l = \overline{Y}_l$. Calculate the value of the objective function in (1) by summing total generalized travel cost of users (obtained by the cross entropy learning algorithm described below) and operation cost.
If $Z(\mathbf{y}', \mathbf{f}^*) < Z(\mathbf{y}_i, \mathbf{f}^*)$, set $\mathbf{y}_i = \mathbf{y}'$ and $Z(\mathbf{y}_i, \mathbf{f}^*) = Z(\mathbf{y}', \mathbf{f}^*)$; otherwise set trial frequency vector $\mathbf{y}'$ by decreasing $y_l$ in $\mathbf{y}_i$ by $\Delta$. If the resulting $y_l - \Delta < \underline{Y}_l$, then set $y_l = \underline{Y}_l$.
Compute $Z(\mathbf{y}', \mathbf{f}^*)$, if $Z(\mathbf{y}', \mathbf{f}^*) < Z(\mathbf{y}_i, \mathbf{f}^*)$, set $\mathbf{y}_i = \mathbf{y}'$ and $Z(\mathbf{y}_i, \mathbf{f}^*) = Z(\mathbf{y}', \mathbf{f}^*)$.
Set $i := i + 1$.

**Step 3:** If no improvement found, set $\Delta := \Delta/2$. If the resulting step size smaller than a small positive value, i.e., $\Delta < \xi$, stop the algorithm; otherwise goto Step 2.

### 4.2 Cross entropy learning algorithm

The cross entropy learning algorithm is designed for solving dynamic multimodal user equilibrium (UE) problems. The algorithm considers the user equilibrium is a rare event to be learned. Based on an iterative procedure, the proposed algorithm adaptively learns optimal travel choice probability by minimizing the Kullback-Liebler relative entropy between two consecutive probability distributions. The resulting probability updates shift the users to cheaper choice alternatives towards the UE. The reader is referred to [7][8] for more detailed description. In current application, the user's decision choice concerns only the departure time choice and path choice in the dynamic capacitated transit network.

Consider the users are located at origins aiming to arrive to respective destinations within the desired arrival time $\tau$, assumed the same for all travelers for simplicity. Based on the bounded-rationality assumption, each traveler is assumed to choose a departure time and path following related choice probability distributions $p_h$ and $p_r$. Based on the experienced generalized travel cost $C_r(t, \mathbf{f})$, the optimal choice probabilities towards the UE are iteratively derived. The detail of the cross entropy learning algorithm is described as follows.

**Step 1**: Initialize uniform probability distributions for departure time choice and path choice. Set $p_h = 1/|H|, \forall h \in H$ and $p_r = 1/|R_k|, \forall r \in R_k$. $R_k$ is the path set of OD pair $k$.

**Step 2**: Dynamic transit system simulation and the travel cost calculation. The user agents move into the network according to his/her departure time and path choice. When arriving at his/her destination, compute the experienced generalized travel cost by (10)-(11).

**Step 3**: Update the departure time choice probability by

$$p_h^{w+1} = p_h^w \frac{e^{-\overline{C}_h^w / \gamma^w}}{\sum_{h' \in H} p_{h'}^w e^{-\overline{C}_{h'}^w / \gamma^w}}, \quad \forall h \in H \qquad (12)$$

, where $\overline{C}_h^w$ is the average generalized travel cost with respect to the users choosing the departure time interval $h$ at iteration $w$. $\gamma^w$ is the control parameter resulting from the solution of the following minimization problem:

$$\text{Min } \gamma^w \text{ subject to } \sum_{h \in H} | p_h^{w+1} - p_h^w | \leq \alpha^w \qquad (13)$$

, where $\alpha^w = \kappa / w$ is a numerical divergent series such that the flow adjustment converges. $\kappa$ is a positive constant. $w$ is an iteration index.

**Step 4**: Update the path choice probability according to the average performance of path choice samples by applying the formulas (12)-(13). Set $w := w+1$.

**Step 5**: When $w = w^{\max}$ or the resulting probability updates stabilize, stop; otherwise goto Step 2.

### 4.3. Method of Successive Averages (MSA)

The method of successive averages approach has been widely applied for solving the simulation-based dynamic traffic assignment problem [11][18][19]. The MSA method is a general iterative path flow adjustment scheme for solving fixed point problems. The adjustment process consists of shifting iteratively travelers to cheaper routes. Given known OD demand, the travelers are initially loaded on the time-dependent shortest paths based on free flow travel time. The shortest paths are then iteratively updated based on travelers' experienced travel cost. By shifting travelers to current found shortest paths, user equilibrium can be approximately achieved by an iterative adjustment process. The algorithm is terminated when the gap function converges to a small value or the maximum iteration is achieved.

Existing applications of the MSA method treat only the path choice problem, given known time-dependent demand [11][19]. As we aim at solving the dynamic user equilibrium problem with respect to departure time and path choice, a MSA-based solution scheme is proposed as follows.

**Step 1**: Initialization. Compute the time-dependent shortest paths for each OD pair and assign OD demand on departure time intervals. Set iteration index $w=0$.
a) Generate the shortest path $u_{hk}^0$ for OD pair $k$ and departure time interval $h$ as

$$u_{hk}^0 = \arg \min_{r \in R_{hk}} [\alpha(\pi_r^w + \pi_r^v) + n_r + \Gamma_r], \quad \forall h, k \qquad (14)$$

Initialize the shortest path set $U_{hk}^0 = \{u_{hk}^0\}$ for all $h$ and $k$.

b) Departure time assignment of travel demand $D_k$. Estimate the generalized travel cost $C_{hk}^0$ on the shortest path $u_{hk}^0$ by

$$C_{hk}^0 = \alpha(\pi_{u_{hk}}^w + \pi_{u_{hk}}^v) + n_{u_{hk}}\lambda + \rho_{u_{hk}}(t,\mathbf{f}) + \Gamma_{u_{hk}}, \quad \forall h,k, \tag{15}$$

where $\rho_{u_{hk}}(t,\mathbf{f})$ is the early/late arrival penalty on the shortest path $u_{hk}^0$ when departing at time $t = t_0 + h\Delta$. For each OD pair $k$, sorting the departure time index set $H$ in a ascending way with respect to $C_{hk}^0$. The obtained ascending departure time choice set for OD pair $k$ is denoted as $H_k'$. Assign uniformly the travel demand $D_k$ on the departure time intervals $1,2,...,s$ in $H_k'$, denoted as $\widetilde{H}_k^0$, such that $sQ_{u_{hk}} \leq D_k < (s+1)Q_{u_{hk}}$, where $Q_{u_{hk}}$ is the maximum allowed passenger flow (i.e. number of users transported per departure time interval for a given OD) on the path $u_{hk}^0$ for one departure time interval $\Delta$. The obtained time-dependent demand for $h$ and $k$ is $d_{hk}^0 = \frac{D_k}{s}, \forall h \in \widetilde{H}_k^0$, and 0 otherwise. This assignment makes users utilize the lowest travel cost departure time intervals under path flow capacity constraints.

**Step 2**: Dynamic network loading. Load all passengers on the network based on their departure time and path choice and run the simulation until all passengers arrive at their destination. Compute the generalized travel cost for all passengers and the value of the gap function.

**Step 3**: If the gap function value is stabilized or the maximum iteration is achieved then stop; otherwise, goto Step 4.

**Step 4**: Update time-dependent link travel time for all used links and compute new time-dependent shortest paths $u_{hk}^{w+1}$ based on Dikjstra's algorithm for each $h$ and $k$.

**Step 5**: Update time-dependent demand $d_{hk}^{w+1}$. Compute average generalized cost for all departure time intervals. Find the least average cost departure time interval $\widetilde{h}_k^{w+1}$. If $\widetilde{h}_k^{w+1} \notin \widetilde{H}_k^w$, then updated time-dependent demand is determined by

$$d_{hk}^{w+1} = \begin{cases} \dfrac{w}{w+1}d_{hk}^w, & \text{if } h \in \widetilde{H}_k^w \\ \dfrac{1}{w+1}D_k, & \text{if } h = \widetilde{h}_k^{w+1} \end{cases} \tag{16}$$

However, if $\widetilde{h}_k^{w+1} \in \widetilde{H}_k^w$, update time-dependent demand by

$$d_{hk}^{w+1} = \begin{cases} \dfrac{w}{w+1}d_{hk}^w, & \text{if } h \neq \widetilde{h}_k^{w+1} \\ \dfrac{w}{w+1}d_{hk}^w + \dfrac{1}{w+1}D_k, & \text{if } h = \widetilde{h}_k^{w+1} \end{cases} \tag{17}$$

**Step 6**: Update path flow assignment for $h$, $k$, $r$ $f_{hkr}^{w+1}$ based on similar formula of (16) and (17). Set $w := w+1$, and goto Step 2.

## 5. NUMERICAL STUDY

In this section, we present and validate the solutions obtained for the lower-level problem by the CE learning algorithm and the MSA algorithm. Then we report the obtained solution of the bilevel problem based on the hybrid algorithm.

The simulation of the transit system is based on the discrete event simulation technique implemented in C++ on a Dell Latitude E6400 with 2.53GHz and 3.48G memory. The proposed algorithm is tested on the multilayer Sioux Falls transit networks (146 nodes and 446 arcs in a multilevel directed graph) by extending LeBlanc's network in [2] (24 nodes and 76 links) (Fig. 2). There are three tramway lines (1, 2, 3) and two metro lines (A and B) with service runs in both directions. The length of arcs is shown in the square brackets of Fig. 2.

The global parameter setting for the experiment is described as follows. The transit modes contain only tramway and metro with strict capacity constraints. The capacity per vehicle for (tramway, metro) are set as (300, 600) and (250, 300) for low and high congestion scenarios. The speed for tramway and metro is set respectively as 5.0 and 12.5 m/sec. The stop time at metro and tramway line nodes is set as 20 seconds for all the vehicles. The walking speed is set as 1.4 m/sec. The length of the boarding, alighting and transfer arcs (between two different stations) is 100 m. The transfer arc from the O/D to connected station node is 300m. For simplicity, the desired arrival time to destination is uniformly set as 9:00. The departure time choice range is set between 7:00 and 9:00 with discretized time interval of 5 minutes. For simplification, $\Gamma_r$ and $\lambda$ are set as 0. The detail of the parameter settings is listed in Table 1.

**Table 1 The parameter settings of the experiments**

| Eq. | Parameter | Value | Eq. | Parameter | Value |
| --- | --- | --- | --- | --- | --- |
| (2) | $\underline{Y}_l$ [1] | 1 | (10) | $\varpi$ | 300 sec. |
| (2) | $\overline{Y}_l$ | 20 | Below (12) | $\kappa$ | 1.6 |
| (1) | $\theta_{l,\text{tramway}}$ | 100 | (1) | $\theta_{l,\text{metro}}$ | 400 |
| (9) | $\alpha$ | 7 | (9) | $\Gamma_r$, $\lambda$ | 0 |
| (10) | $\mu_a$ | 4 | (10) | $\mu_b$ | 15 |
|  | $\mathbf{y}_0$ | {10,10,10,10,10} | (10) | $\tau$ | 9:00 |

Remark: 1. vehicle/hour

### 5.1. Results for the CE learning algorithm and the MSA approach

Before solving the TNDP, we illustrate the performance of the CE learning algorithm and the MSA algorithm for the lower-level problem. The algorithms are tested on different demand level (1600, 4000 and 8000 passengers), vehicle capacity (low and high) and service frequency (2 minutes and 10 minutes). Table 2 shows the computational results of the algorithms. It indicates that at higher congestion level (demand=8000, frequency=10 minutes), the gap function converges to a higher level. This is due to insufficient transit service which makes travelers leave their home quite early and generate higher early arrival penalty. As for the performance of the algorithms, the gap function quickly converges to a stable value after 10 iterations for the CE algorithm. When compared with the MSA method, the CE learning algorithm has similar performance in terms of solution quality and computational times. The validation of obtained user equilibrium solution for the lower-level problem is reported in Table 3, where for each 5 minute departure time interval, the number of users and the average generalized travel cost on the *k*-shortest (k=5) paths are presented. The result indicates that the generalized travel costs on all used paths are no more than that on all unused paths except few exceptions. Note that the generalized travel cost for unused paths on some departure time intervals is estimated by summing in-vehicle travel time, average waiting time (half headway between vehicles) and arrival penalty when departing at the middle point of the departure time interval.

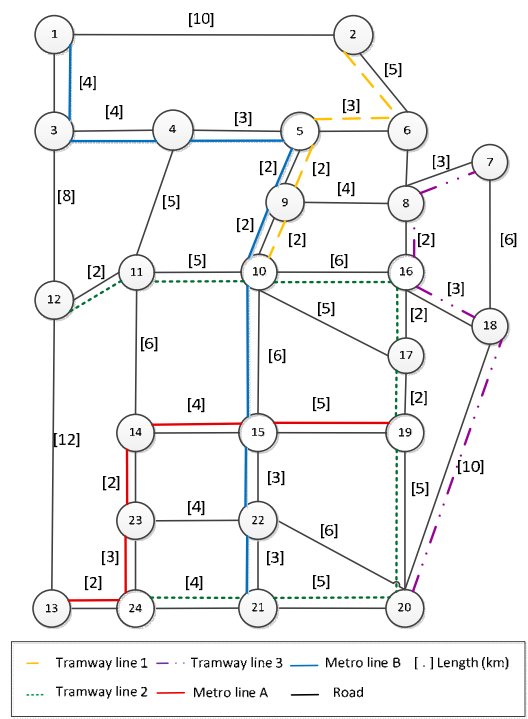

Figure 2 The Sioux Falls network with transit lines

**Table 2 Comparative study of CE and MSA methods for lower-level problem**

| | Demand | | | | | |
|---|---|---|---|---|---|---|
| | 1600 (400 per OD) | | 4000 (1000 per OD) | | 8000 (2000 per OD) | |
| | CE | MSA | CE | MSA | CE | MSA |
| Value of the gap function[1] | 3.03 | 3.00 | 3.03 | 3.05 | 3.07 | 3.14 |
| Total generalized cost | 11878.4 | 11783.2 | 29695.5 | 29811.7 | 60722.0 | 61800.9 |
| Computational time (sec.) | 122.3 | 129.3 | 301.6 | 289.2 | 681.0 | 628.0 |
| Value of the gap function[2] | 4.00 | 4.19 | 7.17 | 7.05 | 14.29 | 14.14 |
| Total generalized cost | 15185.7 | 15872.5 | 82565.3 | 80066.0 | 335274.0 | 331808.0 |
| Computational time (sec.) | 110.0 | 106.6 | 284.3 | 280.9 | 641.2 | 640.0 |

Remarks:
1. Frequency = 2 minutes, capacity for metro and tramway: 600 passengers/veh and 500 passengers/veh, respectively.
2. Frequency = 6 minutes, capacity for metro and tramway: 300 passengers/veh and 250 passengers/veh, respectively.
3 The runtime for 20 iterations.

**Table 3 Validation of obtained solution based on the CE method (total demand = 8000 (2000 for each OD pair), with frequency = 2 minutes for metro and tramway, capacity for metro and tramway : 600 passengers/veh and 500 passengers/veh, respectively)**

| | OD = (1,13) | | | | | | | | | |
|---|---|---|---|---|---|---|---|---|---|---|
| | Number of passengers on each path | | | | | Average generalized travel cost on each path | | | | |
| Departure time interval | r1 | r2 | r3 | r4 | r5 | r1 | r2 | r3 | r4 | r5 |
| 7:25-7:30 | 0 | 0 | 0 | 0 | 0 | 9.11 | 10.14 | 10.14 | 10.27 | 10.53 |
| 7:30-7:35 | **3** | 0 | 0 | 0 | 0 | **8.70** | 10.06 | 10.06 | 10.27 | 10.53 |
| 7:35-7:40 | **22** | 0 | 0 | 0 | 0 | **8.41** | 10.06 | 10.06 | 10.40 | 11.21 |
| 7:40-7:45 | **117** | 0 | 0 | 0 | 0 | **8.07** | 10.99 | 10.99 | 11.65 | 12.46 |
| 7:45-7:50 | **512** | 0 | 0 | 0 | 0 | **7.75** | 12.24 | 12.24 | 12.90 | 13.71 |
| 7:50-7:55 | **660** | 0 | 0 | 0 | 0 | **7.59** | 13.49 | 13.49 | 14.15 | 14.96 |
| 7:55-8:00 | **674** | 0 | 0 | 0 | 0 | **7.57** | 14.74 | 14.74 | 15.40 | 16.21 |
| 8:00-8:05 | **12** | 0 | 0 | 0 | 0 | **8.26** | 15.99 | 15.99 | 16.65 | 17.46 |
| 8:05-8:10 | 0 | 0 | 0 | 0 | 0 | 9.67 | 17.24 | 17.24 | 17.90 | 18.71 |

| | Number of passengers on each path ||||| Average generalized travel cost on each path |||||
|---|---|---|---|---|---|---|---|---|---|---|
| Departure time interval | r1 | r2 | r3 | r4 | r5 | r1 | r2 | r3 | r4 | r5 |
| **OD = (1,20)** ||||||||||| 
| 7:15-7:20 | 0 | 0 | 0 | 0 | 0 | 10.31 | 10.40 | 11.04 | 11.17 | 11.34 |
| 7:20-7:25 | **1** | 0 | 0 | 0 | 0 | **10.24** | 10.07 | 10.71 | 10.90 | 11.29 |
| 7:25-7:30 | **22** | **4** | 0 | 0 | 0 | **9.74** | **9.91** | 10.59 | 10.90 | 11.29 |
| 7:30-7:35 | **128** | **6** | 0 | 0 | 0 | **9.42** | **9.61** | 10.59 | 11.14 | 12.36 |
| 7:35-7:40 | **988** | **12** | 0 | 0 | 0 | **9.15** | **9.50** | 11.42 | 12.39 | 13.61 |
| 7:40-7:45 | **591** | **126** | 0 | 0 | 0 | **9.13** | **9.58** | 12.67 | 13.64 | 14.86 |
| 7:45-7:50 | **122** | 0 | 0 | 0 | 0 | **9.31** | 9.25 | 13.92 | 14.89 | 16.11 |
| 7:50-7:55 | 0 | 0 | 0 | 0 | 0 | 9.78 | 10.50 | 15.17 | 16.14 | 17.36 |
| **OD = (2,13)** ||||||||||| 
| 7:40-7:45 | 0 | 0 | 0 | 0 | 0 | 7.34 | 7.96 | 8.16 | 8.37 | 8.47 |
| 7:45-7:50 | **1** | 0 | 0 | 0 | 0 | **7.00** | 7.62 | 7.82 | 8.26 | 8.47 |
| 7:50-7:55 | **9** | 0 | 0 | 0 | 0 | **6.67** | 7.29 | 7.76 | 8.26 | 8.50 |
| 7:55-8:00 | **52** | 0 | 0 | 0 | 0 | **6.41** | 7.29 | 7.76 | 9.09 | 9.75 |
| 8:00-8:05 | **287** | 0 | 0 | 0 | 0 | **6.01** | 7.29 | 8.75 | 10.34 | 11.00 |
| 8:05-8:10 | **853** | 0 | 0 | 0 | 0 | **5.70** | 8.53 | 10.00 | 11.59 | 12.25 |
| 8:10-8:15 | **727** | 0 | 0 | 0 | 0 | **5.70** | 9.78 | 11.25 | 12.84 | 13.50 |
| 8:15-8:20 | **71** | 0 | 0 | 0 | 0 | **5.98** | 11.03 | 12.50 | 14.09 | 14.75 |
| 8:20-8:25 | 0 | 0 | 0 | 0 | 0 | 7.76 | 12.28 | 13.75 | 15.34 | 16.00 |

|  | OD = (2,20) | | | | | | | | | |
|---|---|---|---|---|---|---|---|---|---|---|
|  | Number of passengers on each path | | | | | Average generalized travel cost on each path | | | | |
| Departure time interval | r1 | r2 | r3 | r4 | r5 | r1 | r2 | r3 | r4 | r5 |
| 7:30-7:35 | 0 | 0 | 0 | 0 | 0 | 8.53 | 8.63 | 9.15 | 9.25 | 9.27 |
| 7:35-7:40 | **2** | 0 | **1** | 0 | 0 | **8.32** | 8.30 | **9.00** | 8.92 | 8.94 |
| 7:40-7:45 | **16** | **3** | 0 | 0 | 0 | **7.92** | **8.01** | 8.52 | 8.75 | 8.80 |
| 7:45-7:50 | **188** | **9** | 0 | 0 | 0 | **7.58** | **7.75** | 8.52 | 8.75 | 8.80 |
| 7:50-7:55 | **753** | **236** | 0 | 0 | 0 | **7.42** | **7.74** | 8.65 | 9.36 | 9.52 |
| 7:55-8:00 | **668** | **23** | 0 | 0 | 0 | **7.58** | **8.21** | 9.90 | 10.61 | 10.77 |
| 8:00-8:05 | **101** | 0 | 0 | 0 | 0 | **7.61** | 7.34 | 11.15 | 11.86 | 12.02 |
| 8:05-8:10 | 0 | 0 | 0 | 0 | 0 | 7.88 | 8.59 | 12.40 | 13.11 | 13.27 |

### 5.2. Results for the hybrid algorithm

The performance of the Hooke-Jeeves algorithm for solving the upper-level problem is shown in Table 4. Two scenarios with respect to different levels of demand are set as 1600 and 8000 users. As can be seen in Table 4, the Hooke-Jeeves algorithm converges efficiently to near-optimal line frequency. Note that we utilize 10 iterations to obtain near user-optimal flow by the CE learning algorithm for solving the lower-level problem in order to reduce the computational time.

## 6. CONCLUSION

In this work, a hybrid multiagent learning algorithm is proposed to solve the dynamic simulation-based transit network design problem. The problem is formulated as a bilevel programming problem where the upper-level is a constrained minimization problem for the optimal transit line frequency decision, and the lower-level is a variational inequality (VI) problem for solving user-optimal equilibrium flow problem. The proposed hybrid algorithm is composed of two main steps to iteratively solve the bilevel problem. In the first step, the cross entropy learning algorithm is proposed to solve the lower-level problem. Then the Hooke-Jeeves algorithm is applied to iteratively find optimal line frequencies for the upper-level problem. Computational results on the extended Sioux Falls network illustrate that the proposed method can find near-optimal solution under dynamic user equilibrium constraints. We compare the cross entropy learning algorithm with the method of successive average for solving dynamic transit assignment problem. The results show that the performance of the two approaches is similar for the test network. Future extensions include applying the queuing theory for modeling passenger flow at stations, and modeling the heterogeneity of passenger's route choice behavior.

**Table 4 Summary of Hooke-Jeeves iterations**

| Total demand= 1600 (400 per OD) | | | | | | | Total demand = 8000 (2000 per OD) | | | | | | |
|---|---|---|---|---|---|---|---|---|---|---|---|---|---|
| Δ | $f_1$ | $f_2$ | $f_3$ | $f_A$ | $f_B$ | Z | Δ | $f_1$ | $f_2$ | $f_3$ | $f_A$ | $f_B$ | Z |
| -- | 10 | 5 | 5 | 5 | 10 | 22834.3 | -- | 10 | 10 | 10 | 10 | 10 | 102850.0 |
| 4 | 6 | 5 | 1 | 5 | 6 | 20724.9 | 4 | 10 | 10 | 10 | 10 | 14 | 92728.5 |
| 4 | 6 | 5 | 1 | 5 | 6 | 20724.9 | 4 | 10 | 14 | 10 | 10 | 18 | 87048.5 |
| 2 | 4 | 5 | 1 | 3 | 6 | 20251.3 | 4 | 10 | 14 | 6 | 10 | 20 | 86669.1 |
| 2 | 2 | 5 | 1 | 3 | 6 | 20186.1 | 4 | 14 | 18 | 6 | 10 | 20 | 84558.6 |
| 2 | 2 | 5 | 1 | 3 | 6 | 20186.1 | 4 | 10 | 20 | 2 | 10 | 20 | 83761.2 |
| 1 | 2 | 5 | 1 | 3 | 6 | 19924.9 | 2 | 10 | 20 | 2 | 10 | 20 | 83761.2 |
| - | - | - | - | - | - | - | 1 | 10 | 20 | 1 | 10 | 20 | 83617.7 |

Remark:

1. $f_1$: tramway line 1; $f_2$ tramway line 2, $f_3$ tramway line 3; $f_A$ metro line A, $f_B$ metro line B.

2. The vehicle capacity setting is 600 passengers and 300 passengers for metro and tramway, respectively.